\documentclass[epj]{svjour}
\usepackage{graphics}
\usepackage{epsf}
\begin{document}
\title{Photoproduction of $\pi^0\eta$ on protons and the $\Delta(1700)D_{33}$ resonance}
\author{
V.L.~Kashevarov\inst{1,2}\thanks{\emph{eMail address:} kashev@kph.uni-mainz.de},
A.~Fix\inst{3},
P.~Aguar-Bartolom\'e\inst{1},
L.K.~Akasoy\inst{1},
J.R.M.~Annand\inst{4},
H.J.~Arends\inst{1},
K.~Bantawa\inst{5},
R.~Beck\inst{6},
V.~Bekrenev\inst{7},
H.~Bergh\"auser\inst{8},
B.~Boillat\inst{9},
A.~Braghieri\inst{10},
D.~Branford\inst{11},
W.J.~Briscoe\inst{12},
J.~Brudvik\inst{13},
S.~Cherepnya\inst{2},
E.J.~Downie\inst{4,1},
P.~Drexler\inst{8},
L.V.~Fil'kov\inst{2},
D.I.~Glazier\inst{11},
R.~Gregor\inst{8},
E.~Heid\inst{1},
D.~Hornidge\inst{14},
O.~Jahn\inst{1},
T.C.~Jude\inst{11},
A.~Knezevic\inst{15},
R.~Kondratiev\inst{16},
M.~Korolija\inst{15},
M.~Kotulla\inst{8},
A.~Koulbardis\inst{7},
S.~Kruglov\inst{7},
B.~Krusche\inst{9},
V.~Lisin\inst{16},
K.~Livingston\inst{4},
I.J.D.~MacGregor\inst{4},
Y.~Maghrbi\inst{9},
D.M.~Manley\inst{5},
M.~Martinez-Fabregate\inst{1},
J.C.~McGeorge\inst{4},
E.F.~McNicoll\inst{4},
D.~Mekterovic\inst{15},
V.~Metag\inst{8},
S.~Micanovic\inst{15},
B.M.K.~Nefkens\inst{13},
A.~Nikolaev\inst{6},
R.~Novotny\inst{8},
M.~Ostrick\inst{1},
R.O.~Owens\inst{4},
P.~Pedroni\inst{10},
F.~Pheron\inst{9},
A.~Polonski\inst{16},
J.~Robinson\inst{4},
S.N.~Prakhov\inst{13},
G.~Rosner\inst{4},
T.~Rostomyan\inst{9},
S.~Schumann\inst{1},
D.~Sober\inst{17},
A.~Starostin\inst{13},
I.M.~Suarez\inst{13},
I.~Supek\inst{15},
C.M.~Tarbert\inst{11},
M.~Thiel\inst{8},
A.~Thomas\inst{1},
M.~Unverzagt\inst{1},
D.P.~Watts\inst{11},
I.~Zamboni\inst{15},
and F.~Zehr\inst{9}.\\
\begin{center}
(The Crystal Ball at MAMI, TAPS, and A2 Collaborations)
\end{center}
}
\institute{
$^1$ Institut f\"ur Kernphysik, Johannes Gutenberg-Universit\"at Mainz, Mainz, Germany\\
$^2$ Lebedev Physical Institute, Moscow, Russia\\
$^3$ Tomsk Polytechnic University, Tomsk, Russia\\
$^4$ Department of Physics and Astronomy, University of Glasgow, Glasgow, UK\\
$^5$ Kent State University, Kent, OH, USA\\
$^6$ Helmholtz-Institut f\"ur Strahlen- und Kernphysik, Universit\"at Bonn, Bonn, Germany\\
$^7$ Petersburg Nuclear Physics Institute, Gatchina, Russia\\
$^8$ II. Physikalisches Institut, Universit\"at Giessen, Giessen, Germany\\
$^9$ Institut f\"ur Physik, Universit\"at Basel, Basel, Switzerland\\
$^{10}$ INFN Sezione di Pavia, Pavia, Italy\\
$^{11}$ School of Physics, University of Edinburgh, Edinburgh, UK\\
$^{12}$ Center for Nuclear Studies, The George Washington University, Washington, DC, USA\\
$^{13}$ University of California at Los Angeles, Los Angeles, CA, USA\\
$^{14}$ Mount Allison University, Sackville, NB, Canada\\
$^{15}$ Rudjer Boskovic Institute, Zagreb, Croatia\\
$^{16}$ Institute for Nuclear Research, Moscow, Russia\\
$^{17}$ The Catholic University of America, Washington, DC, USA\\
}
\date{Received: date / Revised version: date}

\abstract{Total and differential cross sections for the reaction $\gamma p\to\pi^0\eta p$
have been measured with the Crystal Ball/TAPS detector using the tagged photon facility
at the MAMI C  accelerator in Mainz. In the energy range $E_\gamma=0.95-1.4$ GeV the
reaction is dominated by the excitation and sequential decay of the $\Delta(1700)D_{33}$
resonance. Angular distributions measured with high statistics allow us to determine the
ratio of hadronic decay widths $\Gamma_{\eta\Delta}/\Gamma_{\pi S_{11}}$ and the ratio of
the helicity amplitudes $A_{3/2}/A_{1/2}$ for this resonance.
}
\authorrunning{V.L. Kashevarov, A. Fix {\it et al}.}
\titlerunning{Photoproduction of $\pi^0\eta$ on protons ...}
\maketitle
\section{Introduction}\label{intro}

The photoproduction of multiple-meson
states provides information about nucleon excitations which is complementary to that
extracted from reactions with single-meson final states. The main features of the
baryon spectrum may be successfully reproduced by constituent quark models. However, for
many resonance states the detailed information about their properties, such as
photocouplings, hadronic branching ratios is still limited, and production of
multiple-meson states can provide important insights into baryon spectroscopy.

An analysis of these processes is also believed to shed light on the problem of ``missing''
resonances, which are predicted by quark models but have not been seen in $\pi N$
elastic scattering. A simple explanation of the absence of these states is that they are
weakly coupled to $\pi N$ configuration and, therefore, should mostly contribute to
multiple meson production.

The photoproduction of $\pi^0\eta$ pairs on the proton is quite a new topic in
photo-meson physics. In the pioneering work \cite{Weinheimer,Horn} on this reaction, it
was used to search for sequential decays of higher-mass $\Delta$ states. At lower
energies some results for the total cross section have been obtained at the Laboratory of
Nuclear Science (LNS), Japan \cite{Tohoku}. More recently, cross sections as well as
linear beam asymmetries have been measured at the GRAAL facility at ESRF \cite{Ajaka},
and with the Crystal-Barrel /TAPS detector at ELSA \cite{Gutz,Horn2}.

An analysis of the experimental results of Horn \emph{et al.} and Ajaka \emph{et
al.}\,\cite{Horn,Ajaka} together with the theoretical work of D\"oring \emph{et al.}
\cite{Dor} has shown that in the low-energy region the process is mainly governed by the
excitation of the $\Delta(1700)D_{33}$ resonance, which decays into the $\pi\eta N$ final
state via an intermediate formation of $\eta\Delta(1232)$ or $\pi S_{11}(1535)$
quasi-two-body systems. At higher energies, according to the results of Horn \emph{et
al.} \cite{Horn}, other resonances and the $pa_0(980)$ configuration start to come into
play.

The major part of the $D_{33}$ decay into $\pi\eta N$ seems to proceed through the
$\eta\Delta$ channel. This observation is in agreement with predictions of the dynamical
model of the Valencia group \cite{Dor}. The $\pi S_{11}$ channel may be interpreted
entirely in terms of a final-state interaction in which the nucleon appearing after
$\Delta$ decay interacts with the $\eta$ meson via excitation of the $S_{11}(1535)$
resonance. In this model, the production of $\pi S_{11}$ is a
higher order process in comparison to $\eta\Delta$, which is produced directly via the
$D_{33}\to \eta\Delta$ decay.

The decay of $D_{33}(1700)$ and some other $\Delta$ type baryons into $\eta \Delta$
was calculated in ref.\,\cite{Capstick} in a constituent quark model as well as on the basis of the
chiral coupled-channel approach in \cite{Lutz,Sarkar}. These calculations also predict
quite a strong coupling of several weakly established resonances to the $\pi\eta N$
channel.

In spite of visible progress, a detailed empirical study of $\pi^0\eta$ production
dynamics is still needed. In particular, a partial-wave analysis, or its analog for the
production of two mesons, would be very desirable. Some steps in this direction were made
in refs.\,\cite{Horn,FOT}. In ref.\,\cite{Horn} the reaction $\gamma p\to\pi^0\eta p$ was
included in a multi-channel fit. The authors of ref.\,\cite{FOT} have discussed the
angular distributions of the produced particles on the basis of the assumption, that at
any given energy the  amplitude is dominated by a single resonating partial wave.

In this paper, we present new measurements for $\gamma p\to\pi^0\eta p$ for photon
energies from threshold to $E_\gamma=1.4$ GeV, which were obtained with the Crystal
Ball/TAPS detector system at the MAMI C accelerator facility in Mainz. These data will be
used for the phenomenological analysis of $\pi^0\eta$ photoproduction.
The paper is organized as follows. In Sect.\,\ref{details}, we briefly describe the
experimental setup and outline the method of the data analysis. The results are then
interpreted within the formalism developed in ref.\,\cite{FOT}. The aim of this analysis
is to investigate the simplest possible interpretation in terms of a single resonating
$D_{33}$ partial-wave amplitude. A more refined analysis, including other amplitudes and
background contributions, will be published elsewhere. In our simple approach values for
the ratio of $\eta\Delta$ to $\pi S_{11}$ decay widths of the $\Delta(1700)D_{33}$ and
the ratio of the helicity amplitudes are determined. Finally, in Sect.\,\ref{conclusion},
we close with a summary and conclusions.

\section{Experimental setup and data analysis}\label{details}
The experiment was performed at the MAMI C accelerator in Mainz\,\cite{MAMI} using the
Glasgow-Mainz tagged photon facility\,\cite{TAGGER}. The quasi-monochromatic photon beam
covered the energy range from 617 to 1402 MeV with an intensity of $2\times
10^5$\,$\gamma s^{-1}$ MeV$^{-1}$ at 620 MeV. The average energy resolution was 4 MeV.
\begin{figure}
\begin{center}
\resizebox{0.45\textwidth}{!}{%
\includegraphics{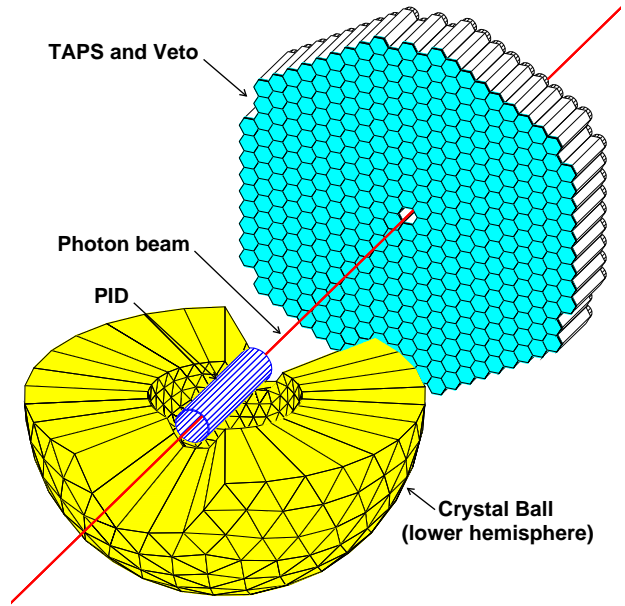}}
\caption{Experimental setup. The upper hemisphere of the Crystal Ball is omitted to show the inside region.}
\label{fig1}
\end{center}
\end{figure}
\begin{figure}
\begin{center}
\resizebox{0.47\textwidth}{!}{%
\includegraphics{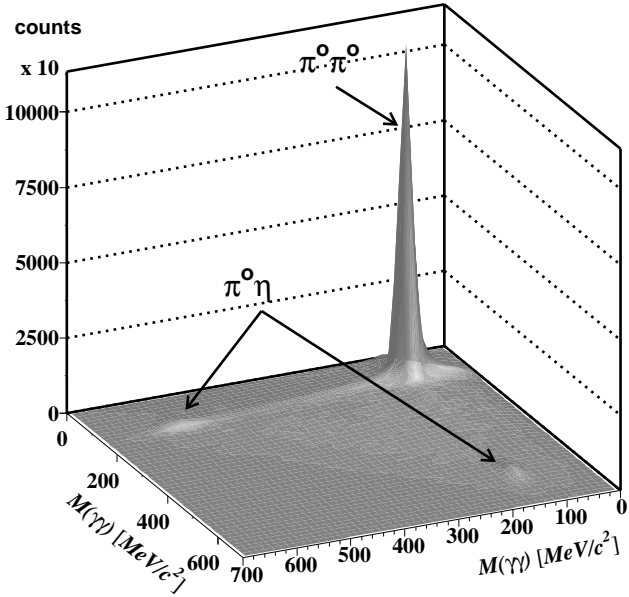}}
\caption{Event selection for final states with 4 photons: $M_{\gamma\gamma}$ {\it vs}
$M_{\gamma\gamma}$ for all possible independent combinations of $\gamma\gamma$ pairs (3
entries for each event).} 
\label{fig2}
\end{center}
\end{figure}

The experimental setup is shown schematically in fig.\,\ref{fig1}. The bremsstrahlung
photons, produced by the electrons in a $10\,\mu$m copper radiator and collimated by a
4-mm-diameter lead collimator, impinged on a liquid hydrogen target with a diameter of
3\,cm and a length of 4.76\,cm. The diameter of the photon beam spot on the target was
about 1 cm. The target was located in the center of the Crystal Ball detector\,\cite{CB}.
This detector consists of 672 optically isolated NaI(Tl) crystals with a thickness of
15.7 radiation lengths covering 93\% of the full solid angle with an energy resolution
for electromagnetic showers of $\Delta E/ E = 1.7\%$ at 1 GeV. Shower directions are
measured with a resolution of $\sigma_{\theta} \approx 2 - 3^\circ$ in the polar and
$\sigma_{\phi} \approx 2^\circ/\sin \theta$ in the azimuthal angle. For charged-particle
identification a barrel of 24 scintillation counters (Particle Identification Detector
\,\cite{PID}) surrounding the target was used.

The forward angular range $\theta = 1 - 20^\circ$ is covered by the TAPS calorimeter
\,\cite{TAPS}. TAPS consists of 384 hexagonally shaped BaF2 detectors, each of which is
25 cm long, which corresponds to 12 radiation lengths. It was installed 147 cm downstream
of the Crystal Ball center. A 5-mm thick plastic scintillator in front of each module
allows the identification of charged particles. The electromagnetic shower energy
resolution of TAPS was $\sigma/E_{\gamma} = 0.0079/(E_{\gamma}/GeV)^{0.5} + 0.018$.
The angular resolution was $0.7^\circ$ (FWHM) for 300 MeV photons. The solid angle of the
combined Crystal Ball and TAPS detection system is nearly $97\%$ of $4\pi$ sr.

The data were collected during two running periods in 2007 (197 and 160 hours,
respectively). In addition, about 70 hours with a double intensity beam were used for a
measurement with an empty target. The trigger threshold for the total energy deposited in
the Crystal Ball detector was 350 MeV.

\begin{figure}[ht]
\begin{center}
\resizebox{0.47\textwidth}{!}{%
\includegraphics{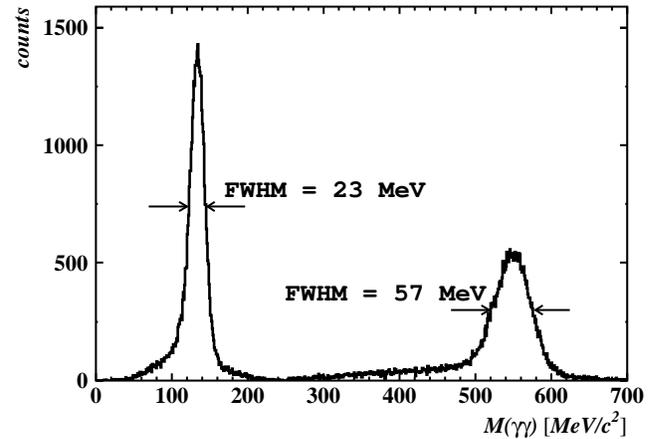}}
\caption{The $\gamma\gamma$ invariant mass spectrum for the best combination of the
$\gamma\gamma$ pairs after $\chi^2$ minimization and rejection of $\pi^0\pi^0$ events.
One pair corresponds to the $\pi^0 \to \gamma  \gamma$ (left), the other to the
$\eta\to\gamma\gamma$ decay (right).} 
\label{fig3}
\end{center}
\end{figure}

In the first step of the identification of the $\gamma p\to \pi^0 \eta p$ reaction,
events with 4 neutral and 1 or 0 charged particles in the Crystal Ball and TAPS detectors
were selected. The $\pi^0$ and $\eta$ mesons were then identified via their decay into 2
photons. The distribution of the invariant masses calculated from
possible $\gamma\gamma$ combinations
is shown in fig.\,\ref{fig2}.
As there are 3 independent combinations for such pairs, this histogram has 3 entries per
event. The distribution shows a large peak corresponding to the $\pi^0\pi^0$ channel and
two smaller ones due to the $\pi^0\eta$ final state. In the next step the $\chi^2$ for
each of the two-meson final states, $\pi^0 \pi^0$ and $\pi^0\eta$, was calculated for the
possible combinations:
\begin{eqnarray}
\chi^2_{2\pi}&=&
\left(\frac{M_{\gamma_i\gamma_j}-m_{\pi^0}}{\sigma_{\pi^0}}\right)^2 +
\left(\frac{M_{\gamma_k\gamma_l}-m_{\pi^0}}{\sigma_{\pi^0}}\right)^2, \label{chi1} \\
\chi^2_{\pi\eta}&=&
\left(\frac{M_{\gamma_i\gamma_j}-m_{\pi^0}}{\sigma_{\pi^0}}\right)^2 +
\left(\frac{M_{\gamma_k\gamma_l}-m_\eta}{\sigma_\eta}\right)^2. \label{chi2}
\end{eqnarray}
Here $m_{\pi^0}$ and $m_\eta$ are $\pi^0$ and $\eta$ masses and $\sigma_{\pi^0} = 10$ MeV
and $\sigma_\eta = 25$ MeV are the corresponding invariant mass resolutions of the
detector system, see fig.\,\ref{fig3}. Each event was now assigned to either $\pi^0\pi^0$
or $\pi^0\eta$ production depending on the minimum of the $\chi^2$ values. After this
selection and a rejection of $\pi^0\pi^0$ events the $\gamma p \to \pi^0\eta p$
reaction can be clearly identified on top of a small background (fig.\,\ref{fig3}). This
histogram has two entries for each event corresponding to the two photon pairs.
\begin{figure*}
\begin{center}
\resizebox{0.75\textwidth}{!}{%
\includegraphics{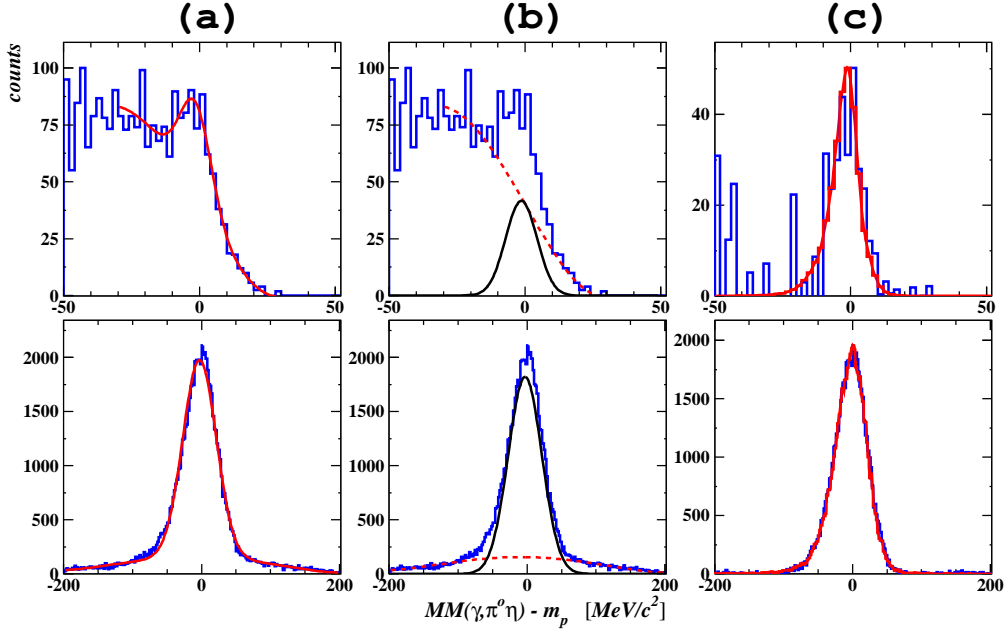}}
\caption{Examples of the background subtraction for lowest ($E_\gamma < 1$ GeV, upper
row) and highest photon energies ($E_{\gamma} > 1.2$ GeV, lower row): (a) experimental
data (histogram) and the best fit by a gaussian + polynomial function (solid line); (b)
the fit components: gaussian (solid line) and polynomial (dashed line); (c) experimental
data after background subtraction (histogram) and GEANT simulation (solid line).}
\label{fig4}
\end{center}
\end{figure*}

After applying a $\chi^2_{\pi\eta}<9$ cut to the two-dimensional $\gamma \gamma$ invariant
mass distribution, the residual background was then eliminated by calculating the missing
mass. In case of a $\gamma p \to \pi^0 \eta p$ reaction the missing mass distribution
calculated from the initial state and the mesons in the final state shows a peak at the
proton mass. Examples of these distributions for the lowest, $E_\gamma<1$ GeV, and the
highest, $E_\gamma>1.3$ GeV, beam energies are shown in fig.\,\ref{fig4}. At lower
energies there is substantial background ($30-60\%$), mainly from the $\gamma p\to
\pi^0\pi^0 p$ reaction that has a three orders of magnitude higher cross section. This
contribution drops rapidly with increasing energy and is reduced to only $\sim 12\%$ for
$E_\gamma>1.2$ GeV. We do not use the data at $E_\gamma<1$ GeV for the calculation of the
angular distributions because of the high background.
\begin{figure}
\begin{center}
\resizebox{0.5\textwidth}{!}{%
\includegraphics{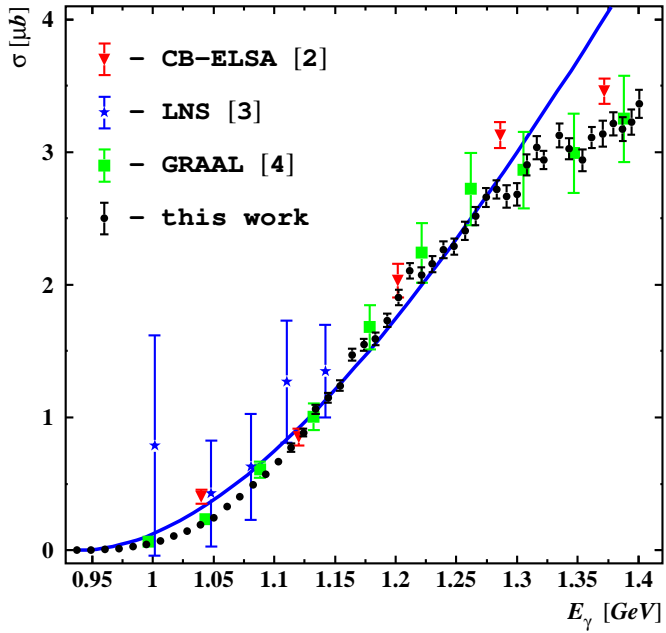}}
\caption{Total cross sections for $\gamma p\to\pi^0\eta p$ reaction. The LNS and GRAAL
points include both statistical and systematic uncertainties. For CB-ELSA and our points
only statistical error bars are shown. The curve shows the energy dependence of the
reaction phase space with arbitrary normalization.} 
\label{fig5}
\end{center}
\end{figure}

The background is subtracted by fitting the missing-mass distributions with the sum of a
gaussian and a third-order polynomial function (fig.\,\ref{fig4} (a) and (b)). After
subtracting the polynomial background, the distribution was found to be in excellent
agreement with results of a Monte Carlo simulation using the GEANT3 code (fig.\,\ref{fig4}(c)).
Finally, to select $\gamma p \to \pi^0 \eta p$ reactions, $3\sigma$
cuts in the missing-mass distributions were applied, and random coincidences with the
tagger, as well as contributions from the target windows, were subtracted. In total,
$\approx 4\times 10^5$ $\gamma p \to\pi^0\eta p$ events were selected. To obtain an
absolute normalization of the cross section, the spectrometer acceptance and the event
reconstruction efficiency were determined using a GEANT3 Monte Carlo simulation. The
average value for the $\pi^0 \eta$ detection efficiency is 20\%. This efficiency includes
branching ratios for $\pi^0$ and $\eta$ decays to two photons.

The photon flux was determined by counting the scattered electrons with the focal-plane
detectors of the tagging spectrometer \cite{TAGGER}. The probability (``tagging efficien- \\
cy'') for a photon to pass through the photon collimator and reach the target per detected
electron was determined with a total-absorption counter that was moved into the beam line at a
reduced photon flux. The tagging efficiency was found to be about 70\% for our experimental
conditions. The systematic uncertainty is estimated to be 5\% and includes uncertainties in
the photon flux, target density and detection efficiency.

\section{Results and discussion}\label{discussion}

\begin{figure*}
\begin{center}
\resizebox{1.0\textwidth}{!}{%
\includegraphics{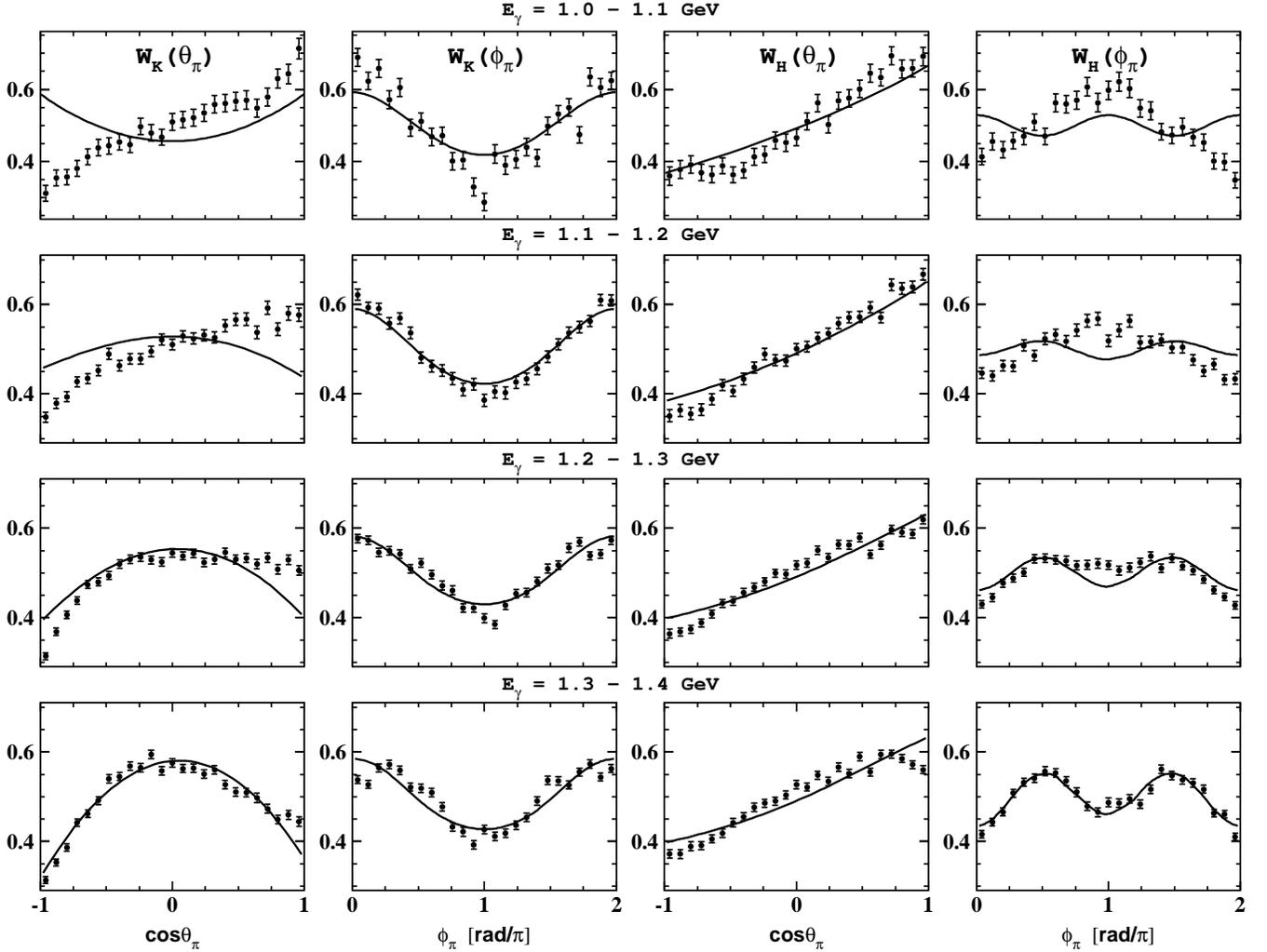}}
\caption{Angular distributions of pions calculated in $\pi p$ c.m.\ frame in the two
coordinate systems explained in fig.\,\protect\ref{fig7}. The data are corrected for the
detector acceptance. Only statistical errors are shown. The notations $W_K$ and $W_H$
are related to the canonical and helicity systems, respectively. The curves represent our model calculation 
which includes only one $D_{33}$ state.} 
\label{fig6}
\end{center}
\end{figure*}
We start the discussion of the results with the total cross section, plotted in
fig.\,\ref{fig5}, where our data (black points) are compared with those obtained at
GRAAL\,\cite{Ajaka}, CB-ELSA\,\cite{Horn} (with statistical errors only) and
LNS\,\cite{Tohoku}. Up to $E_\gamma=1.3$ GeV the cross section exhibits a smooth rise,
reproducing the energy dependence of a three-body phase space (solid line in
fig.\,\ref{fig5}). This suggests that in the low-energy region the reaction mainly
proceeds via formation of $s$-waves in the final system. Otherwise, the centrifugal
barrier appearing in higher partial waves, would suppress the reaction at small relative
momenta. This effect must be especially appreciable near threshold, where the kinetic
energies in the final state are low. As a result, the total cross section would exhibit
more drooping, than that of the phase space. On this basis, it is reasonable to assume
that we are dealing with a resonance, decaying into $\eta\Delta$ in a relative $s$-wave
state. Using spin and parity selection rules one can show that among different partial
amplitudes only $D_{33}$ can produce such a configuration (see, e.g., Table\,I in
ref.\,\cite{FOT}). 

\begin{figure}
\begin{center}
\resizebox{0.45\textwidth}{!}{%
\includegraphics{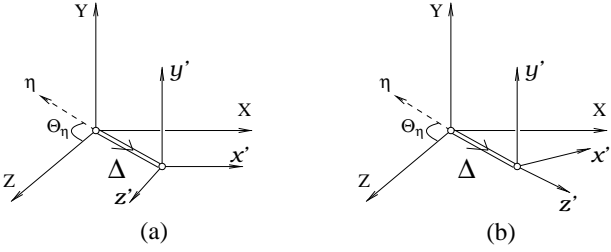}}
\caption{The coordinate systems $(x'y'z')$ used for the analysis of angular distributions
of pions in the $\pi p$ rest frame. In the canonical system (a) the $z'$ axis is 
taken parallel to the beam direction, whereas in the helicity system (b) it is aligned 
along the total $\pi p$ momentum. For both, the $x'$ axis is in the reaction 
plane and the $y'$ axis is chosen as
$\hat{y}'=(\vec{p}_\eta\times\vec{k}_\gamma)/|\vec{p}_\eta\times\vec{k}_\gamma|$.}
\label{fig7}
\end{center}
\end{figure}

The dominance of the $D_{33}$ partial-wave amplitude in the energy range $E_\gamma<1.4$
GeV is confirmed by the angular distributions of the pions in the $\pi p$ rest frame
plotted in fig.\,\ref{fig6}. In order to describe the final-state kinematics, we use the
canonical ($K$) and helicity ($H$) reference frames. Both are fixed to the
$\pi p$ center-of-momentum frame and differ in orientation (see fig.\,\ref{fig7}). The
quantities related to the $K$ and $H$ frames are further denoted by the indices $K$ and
$H$, respectively. The distributions shown in fig.\,\ref{fig6} are normalized by the
total cross section $\sigma_{t}$:
\begin{eqnarray}\label{10}
W(\theta_\pi)&=&\frac{1}{\sigma_t}\int\limits_0^{2\pi}\frac{d\sigma}{d\Omega_\pi}\,d\phi_\pi\,,\\
\label{11}
W(\phi_\pi)&=&\frac{1}{\sigma_t}\int\limits_0^\pi\frac{d\sigma}{d\Omega_\pi}\,\sin\theta_\pi
d\theta_\pi\,,
\end{eqnarray}
where $\theta_\pi$ and $\phi_\pi$ are the pion angles in the corresponding (helicity or
canonical) $\pi p$ center-of-momentum  frame.

\begin{figure}
\begin{center}
\resizebox{0.48\textwidth}{!}{%
\includegraphics{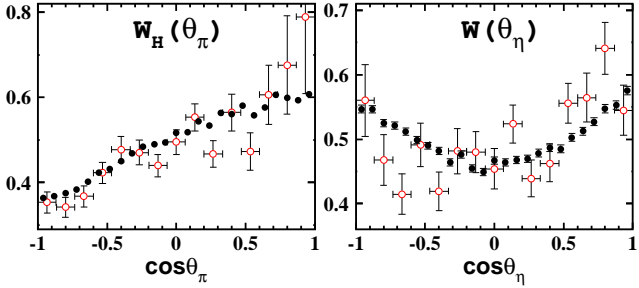}}
\caption{Comparison of our angular distributions (eq.\,\protect(\ref{10})) for $1.72<W<1.87$ GeV (filled circles)to the results from CB-ELSA ref.\,\cite{Horn} for $1.7<W<1.9$ GeV (open circles). 
In both cases the data include statistical errors only. $\theta_{\pi}$ is the angle of the $\pi^0$ 
in the helitiy system and $\theta_{\eta}$ is the angle of the $\eta$ meson with respect to the incoming 
photon in the overall center-of-mass system.}
\label{fig8}
\end{center}
\end{figure}
In fig.\,\ref{fig8} we compare the angular distributions $W_H(\theta_{\pi})$ and $W(\theta_{\eta})$ 
averaged over  $1.1<E_{\gamma}<1.4$ GeV ($1.72<W<1.87$ GeV) to recent results from CB-ELSA \cite{Horn}.
Here $\theta_{\pi}$ is the 
$\pi^0$ angle in the helicity system and $\theta_{\eta}$ the angle of the $\eta$ meson with respect to 
the incoming photon in the overall center-of-mass system.
For comparison, the CB-ELSA data in the energy range $W=1.7-1.9$ GeV (panel (b) in fig.\,9 
of ref.\,\cite{Horn}) were normalized to the integrated cross section. 
The angular distributions in both data sets are in good agreement.

\begin{figure}
\begin{center}
\resizebox{0.45\textwidth}{!}{%
\includegraphics{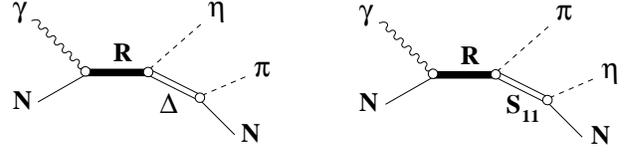}}
\caption{Diagrams representing the amplitude for the $\gamma N\to\pi\eta N$ reaction in a
simple isobar model. The notations $\Delta$ and $S_{11}$ are used for the resonances
$\Delta(1232)P_{33}$ and $N(1535)S_{11}$.}
\label{fig9}
\end{center}
\end{figure}

To gain some insight we calculated the reaction cross section using a simple model which
is similar to the one used in ref.\,\cite{twopion} for double pion photoproduction,
except that we totally neglect the background terms. The latter were calculated in
refs.\,\cite{Dor} and \cite{FOT} and shown to provide only a small fraction of the total
cross section in our energy range. We assume that at a given energy the reaction is
dominated by a single resonating partial wave, $R$. The total reaction amplitude is then
given by a coherent sum of intermediate $R\to \eta\Delta$ and $R\to \pi S_{11}$
transitions represented schematically in fig.\,\ref{fig9}
\begin{equation}\label{15}
t_{m_f\lambda} = A_\lambda(W) \left(F^{(\eta\Delta)}_{m_f\lambda} + F^{(\pi S_{11})}_{m_f\lambda} \right)\,.
\end{equation}
Here the subscripts $m_f=\pm 1/2$ and $\lambda=\pm 1/2,\pm 3/2$ denote the final nucleon
spin projection and the total helicity. The functions $A_\lambda(W)$, depending on the
total c.m.\ energy $W$, are the helicity amplitudes determining the electromagnetic
transition $\gamma N\to R$. The amplitudes $F^{(\eta\Delta)}_{m_f\lambda}$ and $F^{(\pi
S_{11})}_{m_f\lambda}$ describe the decay of the resonance $R$ into $\pi\eta N$ via
intermediate formation of $\eta\Delta$ and $\pi S_{11}$ states. For more details see
\cite{FOT}.

Within the single resonance ansatz (\ref{15}) the angular distributions (\ref{10}) and
(\ref{11}) are determined (apart from the resonance quantum numbers $J^P$) by the ratio
of the partial decay widths,
\begin{equation}\label{25}
r=\left.\frac{\Gamma_{\pi\eta N}^{(\pi S_{11})}}{\Gamma_{\pi\eta N}^{(\eta\Delta)}}\right|_{M_R},
\end{equation}
and the squared ratio of the helicity amplitudes,
\begin{equation}\label{30}
a(W)=\left(\frac{A_{3/2}(W)}{A_{1/2}(W)}\right)^2.
\end{equation}
The two quantities (\ref{25}) and (\ref{30}) were used as adjustable parameters. We
note that $r$ is taken at the resonance position $W=M_R$, so that the energy dependence of
the ratio of $\pi S_{11}$ and $\eta\Delta$ decay widths is fixed by the orbital momenta
associated with these decays. As for the ratio of the helicity amplitudes (\ref{30}),
instead of adopting any parametrization for its energy dependence, we prefer to vary its
values separately in different energy bins.
\begin{figure*}
\begin{center}
\resizebox{0.85\textwidth}{!}{%
\includegraphics{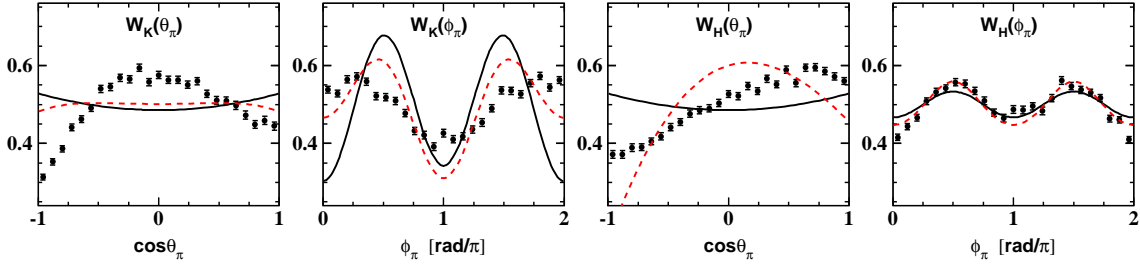}}
\caption{Angular distributions averaged over the energy bin $E_{\gamma}=1.3-1.4$\,GeV.
The curves are calculated with $P_{33}$ (solid curve) and $D_{35}$ (dashed curve)
isobars. In both cases, the parameters were chosen
to give the best description of $W_K(\theta_\pi)$ and $W_H(\phi_\pi)$.}%
\label{fig10}
\end{center}
\end{figure*}

The curves in fig.\,\ref{fig6} are predictions of our simple model containing only
a $D_{33}$ amplitude (that is, $R=D_{33}$ in fig.\,\ref{fig9}).
The results demonstrate that it is possible to get a reasonable agreement with the data
by taking into account only the $D_{33}(1700)$ resonance. A comparable description is
not possible if states with other quantum numbers are used. As an example, in
fig.\,\ref{fig10} the results of analogous calculations with $P_{33}$ and $D_{35}$
amplitudes are plotted. In these cases, the model can at most account for only some of the
distributions, but is unable to reproduce all of them simultaneously.

Other states which were considered in ref.\,\cite{FOT}, namely $S_{31}$, $P_{31}$ and
$F_{35}$, also fail to describe the measured observables. Indeed, as shown in
ref.\,\cite{FOT}, the first two states having $J=1/2$ (and thus $A_{3/2}=0$) lead to an
isotropic $\phi_\pi$ distribution in the helicity frame, i.e. in this case \\
$W^{S_{31}}_H(\phi_\pi)=W^{P_{31}}_H(\phi_\pi)=1/2$, in contrast to our experimental
results. As for $F_{35}$, it has been shown in ref.\,\cite{FOT} that this resonance
should always exhibit a maximum in the distribution $W_H(\phi_{\pi})$ at $\phi_{\pi} =
\pi$ which is not the case in the results in fig.\,\ref{fig6}.
\begin{figure*}
\begin{center}
\resizebox{0.85\textwidth}{!}{%
\includegraphics{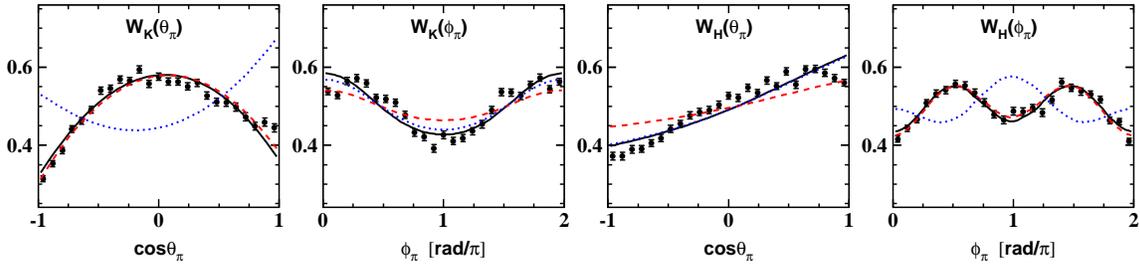}}
\caption{Same as in fig.\,\protect\ref{fig10} but with a $D_{33}$ isobar. The solid curves
are obtained with parameters $r=2/3$\, (see eq.\,\protect\ref{35}) and $a=2.0$\, (see
eq.\,\protect\ref{40a}). They are the same as in the lower panel of
fig.\,\protect\ref{fig6}. The dashed and dotted curves are calculated using the parameter
sets ($r=1/6$, $a=2.0$) and ($r=2/3$, $a=0.7$), respectively. The dotted curve in the
third panel and the dashed curve in the fourth panel  almost coincide with the solid curves.}%
\label{fig11}
\end{center}
\end{figure*}

As shown in fig.\,\ref{fig11}, the distributions
$W_K(\theta_\pi)$ and $W_H(\phi_\pi)$ are sensitive to the ratio $r$
(eq.\,\ref{25}). Even small variations of this parameter
cause significant changes in the shape of $W_H(\theta_\pi)$. At the same time, the other two
distributions $W_H(\phi_\pi)$ and $W_K(\theta_\pi)$ depend strongly on $a$
(eq.\,\ref{30}), whereas the value of $r$ has only little effect.
Clearly, this fact makes the phenomenological analysis of the data easier and allows for
an (almost) independent determination of the values $a$ and $r$.

The curves in fig.\,\ref{fig6} are obtained with
\begin{equation}\label{35}
r=\frac{2}{3}\,.
\end{equation}
We checked that this result is almost independent of the mass $M_R$ used in the
definition (\ref{25}). This value of $r$ is likely to change slightly in a more refined
analysis where other resonances with different spin-parities are included.

If the small influence of the $\pi S_{11}$ decay channel of the $D_{33}(1700)$ is
 neglected, the dependence of $W_H$ on the azimuthal pion angle $\phi_\pi$ has the form \cite{FOT}
\begin{equation}\label{37}
W^{D_{33}}_H(\phi_\pi)=\frac{1}{2\pi}\left(1+\frac{1-a}{3(1+a)}\ \cos 2\phi_\pi\right)\,,
\end{equation}
which shows a  minimum (maximum) at $\phi_\pi=\pi$ for $a>1\,(<1)$. In the region
$E_\gamma=1.2-1.4$ GeV our data exhibit a clear minimum at this point, so $a<1$ can be
excluded. At lower energies the quality of our simple fit becomes worse. In
particular, we find it difficult to describe simultaneously the data for
$W_K(\theta_\pi)$ and $W_H(\phi_\pi)$. This may point to the presence of other resonances
which are not included into our model.

The results presented in fig.\,\ref{fig6} are obtained using
\begin{eqnarray}
a&=&0.7\ \quad \mbox{for}\ E_\gamma=1.0-1.1\ \mbox{GeV}\,,\label{38}\\
& & \nonumber\\
a&=&1.2\ \quad \mbox{for}\ E_\gamma=1.1-1.2\ \mbox{GeV}\,,\label{39}\\
& & \nonumber\\
a&=&1.7\ \quad \mbox{for}\ E_\gamma=1.2-1.3\ \mbox{GeV}\,,\label{40}
\end{eqnarray}
and
\begin{equation}\label{40a}
\!\!\!\!a=2.0\ \quad \mbox{for}\ E_\gamma=1.3-1.4\ \mbox{GeV}\,.
\end{equation}

A direct comparison of the values (\ref{38})-(\ref{40a}) with those given e.g. by the 
PDG\,\cite{PDG} may fail. Our fit of $a(W)$ is energy dependent,
whereas the PDG values of $A_{\lambda}$ are determined at the respective resonance energy
$W=M_R$. As mentioned above, our simple model calculation is not able to decribe the data 
in the energy range from 1670 to 1750 MeV ($E_\gamma=1015-1160$ MeV) given by the PDG  
for the mass of the $D_{33}(1700)$ resonance.  The deviations between the theory and
the data in this region  may indicate onset of other resonances not
included into the present calculation. Therefore, the results (\ref{38} -- \ref{40a}) may
change in a more refined analysis, containing higher partial waves. 
The curve shown in the first row of fig.\,\ref{fig6}
was obtained with a value $a = 0.7$ consistent with the average $a=0.67\pm 0.39$ 
given in the PDG compilation \cite{PDG}.

\begin{figure}
\begin{center}
\resizebox{0.47\textwidth}{!}{%
\includegraphics{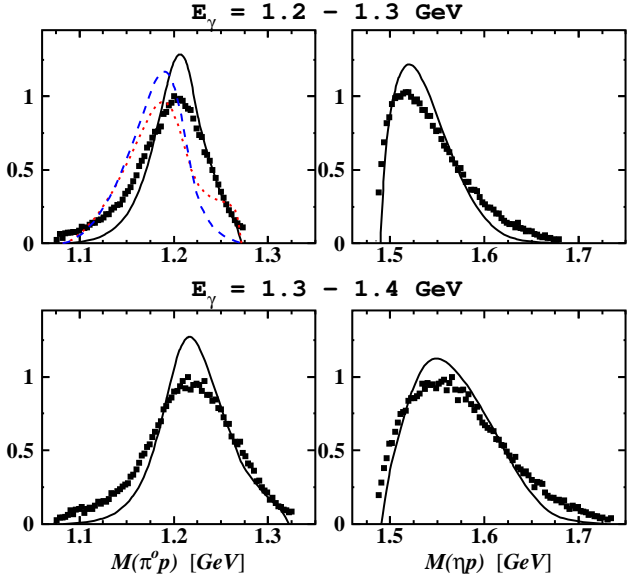}}
\caption{Invariant mass distributions for the $\pi^0 p$ and $\eta p$ subsystems. The
distributions are normalized so that the maximum of the observed distribution is 1. The
data are corrected for the detector acceptance. The solid curves are results of our
isobar-model calculation (eq.\,\protect\ref{15}) with a single $D_{33}$ resonance. The
dashed and dotted curves on the upper left panel are obtained with $P_{33}$ and $D_{35}$
amplitudes, respectively.}
\label{fig12}
\end{center}
\end{figure}

The distributions of $\pi p$ and $\eta p$ invariant masses are presented in
fig.\,\ref{fig12}. The spectra agree rather well with the results of
refs.\,\cite{Horn,Ajaka}. A remarkable feature of the $M_{\pi p}$ distribution is the
maximum close to $W=M_{\Delta}$, indicating the importance of the $\eta\Delta$ mode in
$\gamma p\to\pi^0\eta p$. At the same time, the observed peak is not as pronounced as
that predicted by our single resonance ansatz (\ref{15}) with $R=D_{33}$ (solid curves in
this figure). This observation may point to possible background contributions that are not
included in the model. For instance, the additional background mechanism might be related to
a channel with isospin $T=1/2$, which does not contain the $\eta\Delta$ decay mode and
hence should resemble the phase space in the $M_{\pi p}$ distribution.

In total, taking into account the simplicity of our model (single resonance
with no background terms) the quality of description of the data in Figs.\,\ref{fig6} and
\ref{fig12} is quite satisfactory. In particular, the calculation including
only the
$D_{33}$ partial wave amplitude accounts for the peak position in the $\pi p$ spectrum
(upper left panel in fig.\,\ref{fig12}). As noted above, this resonance is the only
candidate decaying into
$\eta\Delta$ state in a relative $s$-wave. Other resonances providing higher waves
($L\geq 1$) in this configuration, tend to shift the $\Delta$ peak to the lower values of
$M_{\pi p}$. As discussed in ref.\,\cite{FOT}, the shift is caused by the centrifugal
barrier, associated with nonzero angular momentum $L$ of the $\eta\Delta$ decay. As is
shown in the first panel of fig.\,\ref{fig12}, the effect comprises several tens of MeV,
depending on the resonance quantum numbers.
These observations may be considered as additional indication of the $D_{33}$
dominance.

\section{Conclusion and outlook}\label{conclusion}
We have presented the experimental results for the total and differential cross sections
for the reaction $\gamma p\to\pi^0\eta p$. The data were obtained with the
Crystal-Ball/TAPS ca\-lorimeter using the tagged photon facility at MAMI C. The data
for the total cross section agree within given uncertainties with  
previous data from ref.\,\cite{Horn,Tohoku,Ajaka}.
Unlike double pion
photoproduction, the total cross section at energies $E_\gamma\leq 1.4$\,GeV does not
show any pronounced structure and its energy dependence is governed by the smoothly
increasing phase space.

The measured angular distributions are in qualitative agreement with the simplest
calculation in which only the $D_{33}$ partial-wave amplitude is included. This analysis
confirms that in the energy region $E_\gamma<1.4$ GeV the reaction is dominated by the
$D_{33}$ partial wave which can naturally be associated with the resonance $D_{33}(1700)$.
As background contributions are small \cite{Dor,FOT}, $\pi^0\eta$ photoproduction allows
an almost background free study of the $D_{33}(1700)$ baryon.

Deviations of this simple model from the data seen in fig.\,\ref{fig6} and
fig.\,\ref{fig12} may indicate the presence of other resonances whose role should be
taken into account in any refined analysis. In ref.\,\cite{Horn} a significant fraction
of the cross section in our energy region is provided by the $P_{33}(1600)$ and
$P_{11}(1880)$ resonances.

We demonstrated that our data are sensitive to the parameters $a$ (eq.\,\ref{25}) and $r$
(eq.\,\ref{30}) characterizing electromagnetic and hadronic decay properties of the
dominating resonance. In particular, the study of $W_H(\theta_\pi)$ and $W_K(\phi_\pi)$
has shown that $r=2/3$ for $D_{33}$(1700) is
favoured. Furthermore, for the squared ratio of $A_{3/2}$ to $A_{1/2}$ of
$D_{33}(1700)$ we obtain $a=0.7-1.2$ for $E_\gamma=1.0-1.2$ GeV
and $a>1.7$ at higher photon energies.
We would also like to emphasize that our quantitative
results are obtained in quite a simple model and may change in a more sophisticated
approach.

Clearly, the photoproduction on the proton alone does not permit a total determination of
the amplitude, primarily its isospin structure. For the latter purpose, one has to invoke
the reactions on composite nuclear systems, especially on the deuteron and $^3$He, which
are usually used as neutron targets. It should be noted, that not only quasi-free, but
also coherent reactions like $d(\gamma,\pi^0\eta)d$ and
$^3$He$\-(\gamma,\pi^0\eta)^3$\-He, can provide important information. For example, if
our assumption that $\pi^0\eta$ photoproduction is mainly governed by the $T=3/2$ channel
via $\eta\Delta$ is correct, then the amplitudes for proton and neutron should be nearly
equal, and the effect of coherence in reactions of the type $A(\gamma,\pi^0\eta)A$ should
be maximal. In particular, the cross section on the deuteron will be proportional to four
times that on the proton. Significant deviations from this rule would indicate presence
of a $T=1/2$ component in the elementary amplitude. The situation is similar to that
observed in single pion photoproduction in the $\Delta$ region, where dominance of the
$T=3/2$ configuration results in a significant contribution of the coherent channel
$d(\gamma,\pi^0)d$ to the total $d(\gamma,\pi^0)$ rate. Thus, future measurements using
light nuclear targets will help to fully understand the $\pi^0\eta$ photoproduction
amplitude.

\section*{Acknowledgment}
The authors wish to acknowledge the excellent support of the accelerator group and
operators of MAMI. This work was supported by the Deutsche Forschungsgemeinschaft (SFB
443, SFB/TR16), DFG-RFBR (Grant No. 09-02-91330), the European Community-Research
Infrastructure Activity under the FP6 ``Structuring the European Research Area''
programme (Hadron Physics, contract number RII3-CT-2004-506078), Schweizerischer
Nationalfonds, the UK EPSRC and STFC, U.S. DOE, U.S. NSF, and NSERC (Canada). A.F.
acknowledges additional support by the RF Presidential Grant (MD-2772.2007.2). We thank
the undergraduate students of Mount Allison and George Washington Universities for their
assistance.


\end{document}